\documentstyle[preprint,aps,prc,floats,epsfig]{revtex}
\tighten
\let\jnfont=\rm
\def\NPB#1,{{\jnfont Nucl.\ Phys.\ }{\bf B#1},}
\def\PLB#1,{{\jnfont Phys.\ Lett.\ B }{\bf #1},}
\def\PRD#1,{{\jnfont Phys.\ Rev.\ D }{\bf #1},}
\def\PRL#1,{{\jnfont Phys.\ Rev.\ Lett.\ }{\bf #1},}

\def\gsim{\mathrel{\mathpalette\oversim>}}
\def\oversim#1#2{\lower0.5ex\vbox{\baselineskip0pt\lineskip0pt
                 \lineskiplimit0pt\everycr{}\tabskip0pt
                 \halign{$\mathsurround0pt #1\hfil##\hfil$\crcr #2\crcr\sim\crcr}}}
\begin{document}
\draft
\preprint{
\vbox{\hbox{\bf hep-ph/0408271} }}

\title{Late-decaying Q-ball with BBN lifetime }
\author{\ \\[1mm] Fei Wang, Jin Min Yang} 
\address{ \ \\[1mm]
 {\it Institute of Theoretical Physics, Academia Sinica, 
           Beijing 100080, China}}
\maketitle

\vspace*{1cm}
\begin{abstract}
In the Affleck-Dine mechanism of baryogenesis, non-topological solitons called Q-balls can be formed.
In this work we propose that such Q-balls decay during the BBN era and study 
the cosmological consequence of such late decays.   
We find that the late-decaying baryonic Q-balls with lifetime of about $10^3 s$ can provide 
a new developing mechanism for the BBN through a rolling  baryon-to-photon ratio $\eta$, which can 
naturally explain the discrepancy of the BBN prediction with the WMAP data on $^7Li$ abundance. 
For the late-decaying leptonic Q-balls with lifetime of about $10^6 s$, we find that their decay 
product, gravitinos, can serve as a dark matter candidate and 
give an explanation for the approximate equality of dark and baryon 
matter densities. 
\end{abstract}
\vspace*{1cm}


\section{Introduction}
The nature of the matter content of the universe is one of the mysteries in today's physical science.
The Wilkinson Microwave Anisotropy Probe (WMAP) collaboration gives fairly accurate values 
on the contents of our universe \cite{wmap} 
\begin{eqnarray}
       \Omega_m  =  0.27_{-0.04}^{+0.04} \ , ~
       \Omega_b =  0.044_{-0.004}^{+0.004} \ ,~
       \Omega_{\Lambda}=0.73_{-0.04}^{+0.04} \ , ~
       \eta=(6.14\pm0.25)\times10^{-10} \ ,
 \end{eqnarray}
where $\eta$ denotes the baryon-to-photon ratio, and  
$\Omega_m$, $\Omega_b$ and $\Omega_{\Lambda}$ denotes the density of total matter, baryonic matter
and dark energy, respectively.   
One sees that, coincidentally, the dark matter density is comparable to the dark energy density 
as well as to the baryonic matter density. Such coincidences need to be understood. 
While an explanation for the coincidence between dark matter and dark energy can be provided in the quintessence 
scenario, it is hard to give a natural explanation for the coincidence between dark matter and baryonic matter
although some efforts have been devoted \cite{wilczek}. 
A natural explanation for such a coincidence requires some unification scenario  
which correlates the baryongenesis with the dark matter generation.   

The Affleck-Dine (AD) mechanism \cite{affleck} is a promising scenario for baryogenesis.
It is based on the dynamics of a complex scalar field $\phi$ called AD field carrying baryon number. 
In the framework of supersymmetry,  AD field can be realized as a linear combination of 
squark or slepton fields, which has a large expectation value during the 
inflationary stage. The subsequent coherent oscillation creates a large net baryon asymmetry. 
As the coherent oscillation of $\phi$ is generically unstable with spatial perturbation, it 
condenses into non-topological solitons called Q-balls \cite {enqvist}. 
Numerical estimations show that almost all the initial baryon asymmetry of $\phi$ is absorbed 
into these Q-balls.
The baryonic content of the universe can arise from earlier Q-ball evaporation 
which transforms leptonic asymmetry into baryonic asymmetry from sphaleron effects before electroweak 
scale or from the later decay of the baryonic Q-ball. 
At the same time, if the Q-ball is nonstable and
its late decay into LSP (after the freeze out of LSP) will naturally generate the dark matter content 
of the universe \cite{fujii}. 
Note that in the analysis of \cite{fujii}, the Q-balls are assumed to decay before BBN era in order 
not to destruct the BBN predictions for light element abundances. 
However, since the recent WMAP data on light element abundances show some deviation with the BBN 
predictions (as discussed later), it is possible that the Q-balls may decay during the BBN era and 
affect the BBN predictions for light element abundances. 
This possibility was first noted in \cite{ichikawa}.
Such late decays are favorable if their effects on the BBN predictions can improve the fit 
with the observed data. Throughout this work the {\em late-decaying Q-balls} refers to the Q-balls 
which decay during the BBN era.
 
Let us take a look at the BBN predictions for light element abundances compared with the
observed values.  The BBN predictions are given by \cite{bbn-value}
\begin{eqnarray}
&& ~~~~~~ \ {\rm D}/{\rm H} \ < \ 2.49_{-0.18}^{+0.18}\times 10^{-5} \ , \\
&& ~~~~~~~\ Y_p \ = \ 0.249_{-0.001}^{+0.001}\ , \\
&& ~~~~~~~\ ^7{\rm Li}/{\rm H} \ = \  4.7_{-0.8}^{+0.9}\times 10^{-10} \ ,
\end{eqnarray}
while the observed values are given by  \cite{bbn-value}
\begin{eqnarray}
&& 2.40\times 10^{-5} \ < \ {\rm D}/{\rm H} \ < \ 3.22\times 10^{-5} \ , \\
&& 0.227 \ < \ Y_p \ < \ 0.249 \ , \\
&& 9.1\times 10^{-11} \ < \  ^7{\rm Li}/{\rm H} \ < \  1.91 \times 10^{-10} \ ,
\end{eqnarray}
where $Y_p$ denotes $^4$He abundance. 
We see that the $^4He$ and $^7Li$ predictions, especially $^7Li$ prediction,  do not fit well with 
the observed values\footnote{A detailed discussion on this can be found in \cite{bbn2}.}. 

In this work we try to solve the above two problems, i.e., the matter coincidence problem and 
the $^7Li$ abundance problem,  by considering the Q-balls which decay during the BBN era. 
We show that such late-decaying baryonic Q-balls (hereafter called 'B-balls') with lifetime of 
about $10^3 s$  can provide 
a new developing mechanism for the BBN through a rolling $\eta$, which can 
then naturally explain the discrepancy of the BBN prediction with the WMAP data on the light element 
abundances. For the late-decaying leptonic Q-balls (hereafter called 'L-balls') with lifetime of 
about $10^6 s$, we find that their decay products, gravitinos, can serve as a dark matter candidate 
and give a possible explanation of the coincidence between baryonic matter and dark matter. 

\section{ Late-decaying B-Ball and BBN}
\begin{figure}[tb] 
\centerline{\psfig{file=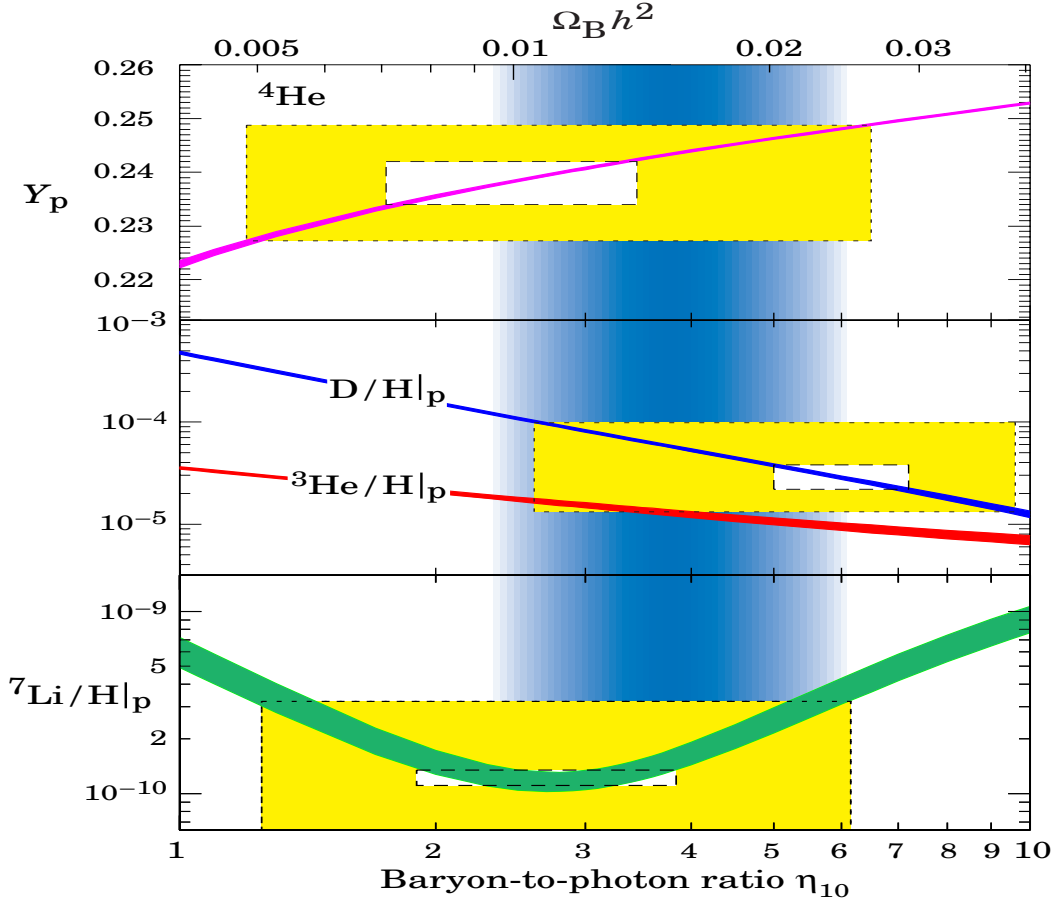,width=14cm,height=12cm,angle=0} }
\vspace*{0.5cm}
\caption{\it The curves are the BBN predictions for light element abundances versus 
          $\eta_{10}(\equiv \eta \times 10^{10})$, taken from Ref.[6]. 
          The boxes denote observed  $1\sigma$ values with and without systematical errors. }         
\label{fig1}
\end{figure}
Fig. 1 shows the BBN predictions for light element abundances versus 
$\eta$ \cite{bbn-value}. We see that, to fit the observed values, BBN predictions 
for $^4He$ and $^7Li$ require a $\eta$ value which is significantly lower than 
$\eta\approx 6\times 10^{-10}$ obtained from the CMB measurement \cite{cmb}.  

In order to solve the $^7Li$ problem,  we propose that 
{\em $\eta$ is rolling and is small during the period of $^7Li$ generation; then $\eta$ rolls up to the ultimate 
observed value after $^7Li$ generation}. 
This mechanism can be realized in the late-decaying B-ball scenario since the decays of B-balls 
can induce the baryonic asymmetry and lead to the rolling of $\eta$ value.
We propose that the BBN undergoes two stages:
\begin{itemize}  
\item[(1)] The first stage is from the start of BBN till the time at which the $^7Li$ synthesis finishes 
           and thus its absolute abundance no longer changes later.
           The baryonic asymmetry arises mainly from the lepton asymmetry through spharleon
           effects instead of from the B-ball decay because the B-ball decay occurs throughout 
           a period which is much longer than the period of this stage.
           This stage is similar to the standard BBN process except that the $\eta$ value is small and is rolling up.  
\item[(2)] The second stage begins when the $^7Li$ synthesis ends while other light elements continue to be
           synthesized \footnote{Due to the continuous injection of the baryonic matter from B-ball decays, 
           the deuterium ($D$) is not much diluted and thus continue to synthesize $^3He$ and $^4He$.}. 
           In this stage, the temperature drops so that the $^7Li$ absolute abundance no longer changes.
           As a result,  the relative abundance of $^7Li$, i.e., $^7Li/H$,  is decreasing since the number 
           of protons is increasing due to the B-ball decays.  
           At the transition point between the first and the second stages,  $\eta$ value is rolling up to 
           $4.21\times 10^{-10}$. We used the FastBBN code \cite{bbn} and adopted three families of neutrinos
           in getting such a $\eta$ value which gives the best fit value of $^7Li$ abundance. 
            We assume that in this stage the ratio of the numbers of protons to neutrons produced from B-ball 
           decays is $p/n \approx 7$ because the baryonic content produced from B-balls reach approximate thermal 
           equilibrium near the surface before they spread out. Taking into account the residual protons from 
           the first stage,   
           the ratio of the total proton number to neutron number should be larger than 7.   
           So the proton consuming reaction is more efficient than in the first stage, which implies that during this 
           stage $^4He$ ($^3He$) synthesis efficiency is relatively lower (higher) than in the first stage.
           Therefore, we predict that the overall synthesis efficiency of $^4He$ ($^3He$) 
           is lower (higher) than the value $0.245$ ($1.3\times10^{-5}$) at $\eta = 4.21\times 10^{-10}$.  
           This means that for $^3He$ our prediction is different from the standard BBN prediction shown in Fig.1.  
\end{itemize} 
The rolling of $\eta$ ends as the B-ball decays finish within the second stage.
Obviously, such a rolling-$\eta$ scenario can give a final required value for the $D$ abundance which
has little difference with standard BBN prediction with the ultimate $\eta$ value ($\approx 6\times 10^{-10}$).      
Therefore, this new BBN generating mechanism through a rolling $\eta$ can not only solve the discrepancy 
between the predicted and observed values of $^7Li$ abundance but also can qualitatively predict the 
required values for $^4He$ and $D$ abundances. 
For the abundance of $^3He$ our scenario predicted a higher value than the standard BBN prediction, 
which remains to be tested by future measurements.

In the following we show that the above scenario can indeed be realized by considering the late-decaying B-balls.

The flat directions  \cite{tony} in the Minimal Supersymmetric Model carry baryon or lepton 
numbers, which can be identified as the AD fields $\Phi$. 
We express the $eLL$ and $udd$ directions as  $\Phi_{L}$ and $\Phi_{B}$.
Although the flat directions are exactly flat when supersymmetry (SUSY) is unbroken, it will be lifted by 
SUSY breaking effects. In the gauge-mediated SUSY breaking (GMSB) scenario, SUSY breaks at low energy scales.  
The shape of the effective potential has the curvature of the order of electroweak scale at low energy and 
almost flat at higher scales \cite{dvali}. 
The gauge mediation part of the potential is given by \cite{kawasaki}
\begin{equation}
  V_{gauge} \sim  \left \{
\begin{array}{ll}
     m_{\phi}^2 {|\Phi|}^2  & ( |\Phi|\ll M_{S} )    \\
     M_{F}^4 (\log\frac{{|\Phi|}^2}{M_{S}^2} ) &  ( |\Phi|\gg M_{S} )
\end{array}
\right.
\end{equation}
where $M_{F}$ is the SUSY breaking scale, $M_{S}$ is the messenger mass scale and $m_{\phi}$ is a soft breaking 
mass of $O$ ( $1 \sim 10$ TeV).

The flat direction can also be lifted by the SUSY breaking mediated by gravity.
The gravity lifted potential is given by 
\begin{equation}
 V_{gravity}\simeq  m_{3/2}^2  \left[  1+ K\log \left( \frac{{|\Phi|}^2}{M_{*}^2} \right) \right] {|\Phi|}^2 \ ,
\end{equation}
where $M_{*}$ is the reduced Planck mass. 
The $K$ term is the one-loop corrections and takes the absolute value $|K|\sim 0.01$ to $0.1$. 
The different sign of $K$ can give different kinds of Q-balls, i.e., a positive $K$ gives 
'delayed type' and a negative $K$ gives 'new type'. 
As noted in \cite{ichikawa}, the contribution to $K$ from Yukawa interactions is positive 
while from gauge interactions is negative.
Since large yukawa couplings arise in the third family,  we can choose $e_2L_1L_3$  flat direction
to give a 'delayed type' L-ball and $u_1d_2d_1$ flat direction to give a 'new type' B-ball \cite{ichikawa}.
This gravity-mediated Q-balls have typical values for their charge and radius \cite{fujii}
\begin{equation}
  Q = Q_{max} \times  \left\{
\begin{array}{lll}
~~\epsilon/\beta     &    ~~~~{\rm for} & ~~~~ \epsilon_{c}\leq \epsilon \leq \beta   \\
~~\epsilon_{c}/\beta &    ~~~~{\rm for} & ~~~~ \epsilon \leq \epsilon_{c}
\end{array}
\right.
\end{equation}
where $\epsilon_{c}\sim 0.01$ and 
\begin{eqnarray}
  &&   Q_{max}\approx 3\times 10^{-3} {\left( \frac{|\phi_{osc}|}{m_{\phi}} \right)}^2 \beta \ ,  \\
  &&   \epsilon\equiv\frac{n_{B} (t_{osc})}{n_{\phi} (t_{osc})} .
\end{eqnarray}
Since the long-life B-ball gives very small $\epsilon$, we have the estimations
\begin{eqnarray}
   Q     & \sim & 3\times 10^{-5} {\left( \frac{|\phi_{osc}|}{m_{\phi}} \right)}^2 \ ,  \\
   R_{Q} & \sim & {(\sqrt{|K|/2}~~ m_{3/2})}^{-1} \ , \\
  \omega & \sim &  m_{3/2} \ .
\end{eqnarray}
The $udd$ flat direction for B-ball can also be lifted by the non-renormalizable terms of the superpotential
given by
\begin{equation}
 W_{NR}^{udd} = \frac{ 9 {(udd)}^2}{ 6 M^3 } =\frac{\Phi_{B}^6}{6 M^3} \ ,
\end{equation}
where the effective scale $M$ includes the coupling $\lambda$ and hence can be larger than the cut-off scale.  
So $\phi_{osc}$ is given as $\phi_{osc}\sim M^{3/4} m_{3/2}^{1/4}$.
The Q-ball decay rate into fermions is proportional to the surface term, given by \cite{cohen}
\begin{equation}
\label{limit}
 \left|\frac{d Q}{d t}\right|\lesssim \frac{{\omega}^3 A}{192 {\pi}^2} \ ,
\end{equation}
where $A$ is the surface of the Q-ball. Here B-balls can decay into nucleons and  
the decay lifetime (in seconds) is given by
\begin{eqnarray}
 \tau_{d}\equiv Q / ( \frac{d Q}{d t} ) = \frac{ 24 \pi \bar{\beta} |K| M^{3/2} }{ m_{3/2}^{5/2} }
= \frac{1.488\times10^{-15} }{ m_{3/2}~Mev} {\left(
\frac{M~/{\rm TeV}}{m_{3/2}/{\rm MeV}}\right)}^{3/2} \ ,
\end{eqnarray}
where $\bar{\beta}=3\times 10^{-5}$.
If we choose the non-renormalizable scale naturally to be reduced Planck scale, it can gives 
a lifetime $10^3 \sim 10^4 s$ for a gravitino a little heavier than 1 GeV. 
This means that the decay of such "new type" B-balls into nucleon can indeed occur during the BBN era. 
Since the B-ball decays can provide baryonic content, the $\eta$ value is natually rolling up.
So the rolling-$\eta$ scenario can be naturally realized in such late-decaying B-ball case.

\section{Late-decaying L-ball and dark matter}
The Q-balls may have very long lifetime in the pure gauge-mediated scenario
and  rather short lifetime in the pure gravity-mediated scenario.
As studied in \cite{kasuya,kawasaki}, in the pure gauge-mediated scenario their lifetime 
can be so long that they may survive decaying and evaporating and exist in today's universe 
as the dark matter. 
In this section we consider the late-decaying 'delayed-type' L-balls with BBN lifetime 
in the general SUSY breaking model with both gravity and gauge effects and study the possibility 
for their decay products to serve as the dark matter. 
We find that this approach is feasible and
can provide an unified scenario to give an explanation for the coincidence
between dark matter and baryonic matter contents.
Note that although such late-decaying L-ball decays occur throughout the BBN era
and affect the BBN predictions for light elements abundances, we find that
the effects can be within the allowed region.

When the zero-temperature potential $V_{gauge}$ dominates at the onset of the coherent oscillations of AD fields,
the gauge-mediated type Q-balls are formed. Their mass $M_{Q}$ and size $R_{Q}$ are given by
\begin{eqnarray}
&&  M_{Q}\sim M_{F} Q^{3/4}\ , \\
&&  R_{Q}\sim M_{F}^{-1} Q^{1/4} \ .
\end{eqnarray}
Just as indicated in the numerical simulations \cite{kasuya}, the produced Q-ball almost absorb all the charges 
carried by the AD field and the typical charge is estimated as
\begin{equation}
  Q \approx \beta {\left( \frac{\phi_{osc}}{M_{F}} \right)}^4 \ ,
\end{equation}
where $\beta \approx 6\times 10^{-4}$.
When $V_{gravition}$ dominates the potential at the onset of the coherent oscillation of the AD fields, 
the 'delayed-type' Q-ball form when the AD field leaves this region and enters the $V_{gauge}$ 
dominated region if $K$ is positive. However, if $K$ is negative, the gravity-mediated type Q-balls ('new-type') 
are produced, as discussed in the preceding section.

So when the AD field starts to oscillate ($H_{osc}\sim \omega \sim m_{3/2}$)
in the $V_{gravition}$ dominated region for positive $K$, the  charge of the 'delayed-type' Q-ball 
is given by
\begin{equation}
 Q \sim \beta {\left( \frac{\phi_{eq}}{M_{F}} \right)}^4 \sim \beta {\left( \frac{M_{F}}{m_{3/2}} \right)}^4 \ .
\end{equation}
Here the gravition term domination condition gives $\phi \geq \phi_{eq} \sim M_{F}^2 / m_{3/2}$.
For L-ball, the upper limit of the decay rate can be estimated as in Eq.(\ref{limit}). 
So the lower limit of the lifetime in seconds is given by 
\begin{equation}
 \tau_{d} \gsim \frac{48 \pi \beta^{1/2} m_{F}^4}{m_{3/2}^5}
         \approx 2431 \times \frac{ { (M_{F}/{\rm TeV}) }^4 } {  {(m_{3/2}/{\rm MeV})}^5} \ .
\end{equation}
\begin{figure}[tb] 
\centerline{\psfig{file=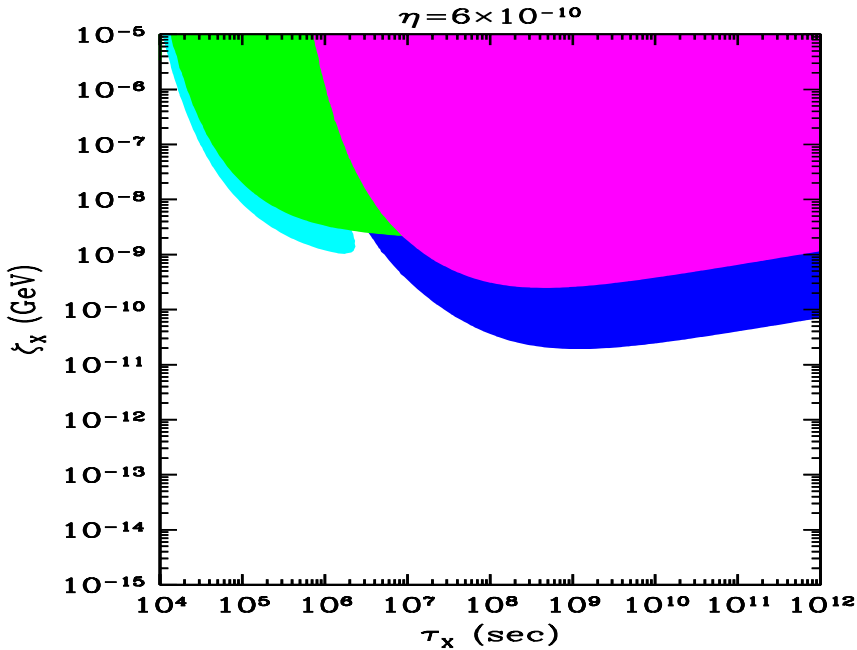,width=12cm,height=10cm,angle=0} }
\vspace*{-0.8cm}
\caption{\it The electromagnetic energy release versus the release time for $\eta=6\times10^{-10}$, 
             taken from Ref.[15]. The shaded regions are excluded by BBN constraints. }     
\label{fig2}
\end{figure}
The late decay of the L-balls will release electromagnetic energy and can alter some light element abundances
since such electromagnetic energy injection can have dissociation effects on light elements.
As shown in Fig. 2 taken from \cite{cyburt}, if such electromagnetic energy release is sufficiently small, 
it will  be allowed.
So if we require that the L-balls keep on decaying until about $10^6 s$ with very small 
energy release, it will not affect the success of BBN. (Note that from Fig. 2 one sees that the
lifetime requirement is not stringent as long as the energy release is sufficiently small. Here 
we require a lifetime of  about $10^6 s$ as an example.)  
We found that such a lifetime requirement can be met by choosing a gravitino mass of $0.24 \sim 1.5$ MeV 
for SUSY breaking scale of $1 \sim 10$ TeV. In the following we show that in this case 
the energy release can be sufficiently small, and the baryonic content and the dark matter content
can be correctly related.   
 
In this case the charge of L-ball can be estimated to be $10^{24}$ and the energy per L-ball charge is 
given by
\begin{equation}
  M_{Q}/Q=M_{F}/Q^{1/4} \sim 10 {\rm ~MeV} .
\end{equation}
It is easy to see that the decay products can only inject electromagnetic energy.
The average electromagnetic energy release for each Q-ball is given by
\begin{equation}
  \zeta_{EM}=\epsilon_{EM} n_{decay} N_{portion} / n_{\gamma}^{BG} \ ,
\end{equation}
where $n_{decay}$ denotes the decaying rate of Q-ball, $\epsilon_{EM}$ is the average electromagnetic 
energy release per AD field decay, $N_{portion}$ is the AD field number generated per Q-charge and it is 
unity here.  Since the tree-level two-body amplitude dominates, the decay mode is $AD \to LSP + S$,
where $LSP$ denotes the lightest supersymmetric particle and $S$ a SM particle.
Since in our analysis gravitino is quite light, much lighter than neutralino, it is assumed
to be the LSP.
So the decay products of the AD fields provide a natural dark matter candidate--gravitino.
The energy of the nearly massless particle $S$ in the decay is given by
\begin{equation}
E_{EM}=\frac{m_{AD}^2-m_{LSP}^2}{2 m_{AD}} \ .
\end{equation}
The energy release rate is given by the decay rate times the energy release per gravitino 
generation 
\begin{equation}
    \zeta_{EM}=\epsilon_{EM}Y_{d}=\epsilon_{EM} \Delta_{Q} n_{Q}  / n_{\gamma}^{BG} \ ,
\end{equation}
where $\Delta_{Q}$ denotes the charge per Q-ball which decayed. 

Suppose some of the initial baryon asymmetry is induced from lepton asymmetry through L-ball evaporation 
with sphaleron effects before the $M_{weak}$ scale, then we have  
\begin{equation}
 \epsilon Q_{i} n_{Q}= f n_{B} ,
\end{equation}
where $Q_{i}$ indicate initial charge per Q-ball, and $n_{Q}$ and $n_{B}$ are the number density for Q-ball
and baryon, respectively. $\epsilon$ denotes the changed portion of lepton asymmetry into baryon asymmetry 
and $f$ be the baryon asymmetry contribution from L-ball to the total asymmetry.

Finally we obtain 
\begin{equation}
 \zeta_{EM}=\frac{\epsilon_{EM} N_{portion} \Delta_{Q} n_{Q}}{n_{\gamma}^{BG} }
           =\frac{\epsilon_{EM} N_{portion} \Delta_{Q} n_{B} f}{n_{\gamma}^{BG} \epsilon ~ Q_{i} }
           =\frac{\epsilon_{EM}  N_{portion} \eta_{B} f }{ \epsilon ~ \tau_{d} } \ .
\end{equation}
As noted in \cite{feng}, the daughter electron will immediately initiate an electromagnetic cascade to
give $\epsilon_{EM}\simeq E_{e}$. For the relativistic daughter muon, it typically 
interacts with the background photons through Thomson scattering. As the scattering time is shorter than 
the time-dilated muon decay time, the muon energy can safely change into electromagnetic energy.
For the daughter tau, the typical scattering life time is longer than the time-dilated tau decay time, 
the electromagnetic energy change ratio is 1/3 to 1. Accounting all the branch ratio of its decay channel 
gives the 1/2 of the 
tau energy as electromagnetic energy release. For simplicity, we chose the release of  electromagnetic
energy to be half of the average lepton energy.

Sufficiently small electromagnetic energy release is possible and, for example, we require 
$\zeta_{EM}\leq 10^{-14}$ GeV, which, combined with the lifetime requirement, is well 
allowed by the BBN constraints shown in Fig. 2.  
Under such requirement we get the approximate portion $\epsilon/f\sim 10^{-4}$.
Then we can give an estimation of the baryonic content of the universe if all the dark matter is 
from L-ball decay
\begin{equation}
 \Omega_{B} /\Omega_{CDM} \sim 10^{-4}\times 1 GeV/ (0.2 \sim 1.5 MeV) \sim O(0.1) \ . 
\end{equation} 
We see that this scenario can give an explanation for the matter coincidence problem. 

\section{conclusion}
We studied the possibility for the Q-balls to decay during the BBN era ( such possibility has not been studied 
in the literature) and the cosmological consequences of such late decays.   
Our findings are:
\begin{itemize}
\item[(1)] The late-decaying baryonic Q-balls with lifetime of about $10^3 s$ can provide 
           a new developing mechanism for the BBN through a rolling baryon-to-photon ratio $\eta$. 
           We proposed that in this scenario the BBN era undergoes two stages separated by    
           a transition point $\eta\approx 4.21 \times 10^{-10}$ and the $^7Li$ synthesis finished   
           in the first stage.
           Such a scenario can not only solve the discrepancy between the predicted and observed 
           values of $^7Li$ abundance, but also can qualitatively predict the required values for 
           $^4He$ and $D$ abundances.  
           For the abundance of $^3He$ this scenario predicted a higher value than the standard BBN, 
           which leaves to be tested by future measurements.
\item[(2)] The late-decaying leptonic Q-balls with lifetime of about $10^6 s$ decay into gravitinos, which
           can serve as a dark matter candidate and give an explanation for the approximate equality of dark 
           and baryon matter densities. 
\end{itemize}

\section*{Acknowlegents}
We are grateful to  Xiaotao He for providing us some useful nuclear data and Junjie Cao for helpful discussions. 
This work is supported in part by National Natural Science Foundation of China.

\end{document}